\documentclass[twocolumn,superscriptaddress,aps,preprintnumbers,amsmath,amssymb,footinbib,prl]{revtex4}
\usepackage{graphicx}
\usepackage{epstopdf}
\usepackage{dcolumn}
\usepackage{bm}
\usepackage{hyperref}
\def\slashchar#1{\setbox0=\hbox{$#1$} 
\dimen0=\wd0 
\setbox1=\hbox{/} \dimen1=\wd1 
\ifdim\dimen0>\dimen1 
\rlap{\hbox to \dimen0{\hfil/\hfil}} 
#1 
\else 
\rlap{\hbox to \dimen1{\hfil$#1$\hfil}} 
/ 
\fi}

\def\d{\delta}

\def\g{\gamma}

\def\l{\lambda}

\def\s{\sigma}

\def\beq{\begin{eqnarray}}
\def\eeq{\end{eqnarray}}

\newcommand{\vev}[1]{ \left\langle {#1} \right\rangle }

\newcommand{\etal}{{\em et al.}}

\newcommand{\vrel}{v_{\text{rel}}}

\begin{document}
\title{Distinguishing Dark Matter Annihilation Enhancement Scenarios via Halo Shapes }

\author{Masahiro Ibe}
\affiliation{%
Department of Physics and Astronomy,
University of California, Irvine, California 92697, USA
}%
\author{Hai-bo Yu}
\affiliation{%
Department of Physics and Astronomy,
University of California, Irvine, California 92697, USA
}%

\begin{abstract}
Sommerfeld enhancement and Breit--Wigner enhancement
of the dark matter annihilation have been proposed to explain
the ``boost factor" which is suggested by observed cosmic ray 
excesses.
Although these two scenarios can provide almost indistinguishable
effects on the cosmic ray fluxes, the cross sections of the self-interaction
in those enhancement mechanisms are drastically different.
As a result, we show that they can be distinguished by examining the effects of the self-interaction on the halo shapes. 
In Sommerfeld enhancement models with $m_\phi\lesssim100~{\rm MeV}$ and $m_{\rm DM}\lesssim 3~{\rm TeV}$, the self-interaction can leave observable imprints in the galactic dynamics, while 
dark matter is effectively collisionless in Breit--Wigner models.
\end{abstract}

\date{\today}
\maketitle
\preprint{}

{\it Introduction}\\
Recent observations of the PAMELA~\cite{Adriani:2008zr},
ATIC~\cite{:2008zz}, PPB-BETS~\cite{Torii:2008xu} and Fermi~\cite{Abdo:2009zk} experiments
strongly suggest the existence of a new source of electron/positron fluxes in cosmic rays. Although the excesses may have astrophysical sources~\cite{Hooper:2008kg,Dado:2009ux}, the 
annihilating dark matter interpretation remains an interesting possibility. 
If dark matter is a thermal relic, however, there is a tension between
the dark matter density and the observed excesses in this interpretation.
That is, the required annihilation cross section of dark matter in the galactic halo is
much larger than the one appropriate to explain the dark matter relic density precisely measured by the WMAP experiment~\cite{Komatsu:2008hk}, i.e. $\vev{\sigma_{\rm ann} v_{\rm rel}}\simeq 3\times 10^{-26}$\,cm$^3/$s,
which we call the WIMP cross section. 
As a result,
the dark matter explanation of the excesses requires
an enhanced annihilation cross section in the galactic halo
by a factor ${\cal O}(10^2-10^3)$ with respect to the WIMP cross section
for dark matter with a mass in the TeV range.

So far, there have been two proposals to explain the boosted annihilation cross section in
particle physics.
The one is the Sommerfeld
enhancement~\cite{Hisano:2004ds,ArkaniHamed:2008qn} and
the other is the Breit--Wigner enhancement~\cite{Ibe:2008ye,Guo:2009aj}.
In the Sommerfeld enhancement scenario,
the dark matter annihilation is enhanced in a low-velocity environment
due to an attractive force among dark matter, which is mediated by a light particle. 
In the Breit--Wigner enhancement scenario,
dark matter annihilates via a narrow Breit--Wigner resonance,
and the cross section is enhanced in a low-velocity environment when
the difference between the resonance mass and the twice of the dark matter
mass is much smaller than the width of the resonance. 
(See Refs.\,\cite{Cirelli:2008pk,Pospelov:2008jd,Feldman:2008xs}
for  recent attempts to attribute the boost factor to the resonance.
See also Refs.\,\cite{Griest:1990kh,Gondolo:1990dk,Jungman:1995df}
for general discussions on the effects of the resonance 
to the dark matter annihilation.)

Since both scenarios were introduced to explain the cosmic ray excesses
in a low-velocity environment, it is rather difficult to distinguish them
by examining cosmic ray fluxes. In this note, we explore the possibility to distinguish them by investigating the morphology of dark matter halos. 

As shown in Refs.~\cite{Feng:2009hw,Buckley:2009in}, in the Sommerfeld enhancement scenario, a light particle 
which enhances the annihilation cross section also mediates self-interaction of dark matter. 
The rather strong self-interaction mediated by the light particle can cause too much energy exchange of dark matter, which leads to spherical shapes of dark matter halos.
As we will discuss below, on the other hand,  self-interaction is highly suppressed 
 in the Breit--Wigner scenario, and its effects on the galactic dynamics are negligible. 
We show that the halo shape effects of the dark matter self-interaction can be used to distinguish two scenarios. 
\\

{\it Dark matter self-interaction in two scenarios}\\
In the Sommerfeld enhancement scenario,
the self-interaction process is dominated by the $t$-channel exchange of the light particle $\phi$, which is inevitable for the this type of model. To illustrate the physical process intuitively, let's look at the differential cross section in the Born approximation,
\begin{eqnarray}
\label{eq:diff}
\frac{d\sigma}{d\Omega} = \frac{\alpha_X^2}
{m_{\rm DM}^2[m_\phi^2/m_{\rm DM}^2+v^2_{\rm rel} \sin^2(\theta_*/2)]^2}\ ,
\end{eqnarray}
where $m_{{\rm DM},\phi}$ are masses of dark matter and the light particle, respectively,
$\alpha_X$ denotes the fine structure constant $\alpha_X = \l^2/(4\pi)$,
and $v_{\rm rel}=\left|\vec{v_1}-\vec{v_2}\right|$ is the relative velocity of dark matter. 
The energy transfer cross section $\sigma_T=\int d\Omega(d\sigma/d\Omega)(1-\cos\theta_*)$ is given by~\cite{Feng:2009hw}
\begin{equation}
\sigma_T = \frac{2 \pi}{m_\phi^2} \beta^2
\left[ \ln  \left( 1 + R^2 \right) - \frac{R^2}{1 + R^2} \right]\ . 
\label{eq:sigmatransfer}
\end{equation}
Here, $\beta \equiv 2 \alpha_X m_\phi/(m_{\rm DM} \vrel^2)$ is a ratio of
the potential energy caused by the light particle at the interaction range, $r\sim m_\phi^{-1}$, 
to the kinetic energy of dark matter, and $R \equiv m_{\rm DM} \vrel / m_\phi$ is the ratio of the
interaction range to the dark matter particle's de Broglie wavelength. 
Notice that  Eq.\,(\ref{eq:sigmatransfer}) receives significant corrections for $\beta\gg1$, 
and we will discuss it in the next section. 
For values of interest here, $\vrel \sim 10^{-3}$
and $m_{\rm DM}/m_\phi \agt 10^3$,  $R$ is typically larger than one. 
Thus, for example, the energy transfer cross section is given by,
$\sigma_T \approx \frac{8 \pi \alpha_X^2}{\vrel^4 m^2_{\rm DM}} \left(\ln R^2
- 1 \right)$
for  $R \gg 1$. 
Although the finite interaction length of the Yukawa potential, $r\sim m_{\phi}^{-1}$, 
cuts off the logarithmic divergence, the energy transfer cross section 
is still greatly enhanced for small $\vrel$.

The annihilation cross section, on the other hand, can be approximated by
\begin{equation}
\sigma_{\rm ann}\sim\frac{\pi\alpha^2_X}{m^2_{\rm DM}\vrel}{\rm min}\left[\frac{\alpha_X}{\vrel},\frac{\alpha_X m_{\rm DM}}{m_\phi}\right].
\end{equation}
The typical value of the enhancement factor ${\rm min}\left[\alpha_X/\vrel,\alpha_X m_{\rm DM}/m_\phi\right]$ is ${\cal O}(10^2)$ for $v_{\rm rel}\sim 10^{-3}$ if we require correct relic abundance of dark matter~\cite{Feng:2009hw,Dent:2009bv}. 
Thus,  the energy transfer cross section $\sigma_T$ is much larger than the annihilation cross section 
$\sigma_{\rm ann}$ by a factor of ${\cal O}(10^7)$ in the case of $R\gg 1$, although both processes are determined by the $t$-channel process. 
This is not surprising  because the annihilation requires a heavy dark matter exchange in the $t$-channel 
and the process is dominated by the $s$-wave, while the scattering is mediated by the light particle $\phi$ 
and higher modes of the partial wave have significant contributions.

Now let us consider the self-interaction in the Breit--Wigner scenario.
In this case,  the annihilation cross section
and the self-interaction are both dominated by the processes of the $s$-channel exchange of
the narrow resonance which can be expressed by the Breit--Wigner forms;
\begin{eqnarray}
&\s_{\rm ann}&  \simeq  \frac{8\pi}{m_{\rm DM}^2\vrel}\frac{\gamma^2}{(\delta+v^2/4)^2+\gamma_s^2}\frac{B_{\rm DM}}{\sqrt{1-4m^2_{\rm DM}/E_{\rm CM}^2}}B_{f}\ ,\cr
&\s_{T}& \simeq  \frac{8\pi}{m_{\rm DM}^2}\frac{\gamma^2}{(\delta+v^2/4)^2+\gamma_s^2}
\left(\frac{B_{\rm DM}}{\sqrt{1-4m^2_{\rm DM}/E_{\rm CM}^2}}
\right)^2.
\label{eq:selfBW}
\end{eqnarray}
Here, $\g_s$ is the total decay width of the resonance $s$ normalized by the resonance mass $m_s$,
$E_{\rm CM}$ is the energy in the center of mass frame,
$B_{\rm DM, f}$ denote the branching ratios of the resonance into a pair of dark matter
and the final state particles, respectively, and $\delta$ is defined by $m_s^2 = 4 m_{\rm DM}^2(1-\d)$ 
with $|\delta|\ll1$.
For simplicity, we assume that the pole is in an unphysical region, i.e. $\delta >0$, 
although our analysis can be extended for a physical pole region straightforwardly (see also discussion in Ref.\,\cite{Ibe:2009en}).  

The ratio of the energy transfer cross section to the annihilation cross section in the Breit--Wigner scenario is
\begin{eqnarray}
\frac{\sigma_T}{\sigma_{\rm ann}}\simeq \frac{B_{\rm DM}}{\sqrt{1-4m^2_{\rm DM}/E_{\rm CM}^2}B_f}\times \vrel\ .
\end{eqnarray}
As discussed in Refs.~\cite{Ibe:2008ye}, 
a successful enhancement factor
can be obtained for
\begin{eqnarray}
\label{eq:BWcond1}
\frac{B_{\rm DM}}{\sqrt{1-4m^2_{\rm DM}/E_{\rm CM}^2}}\lesssim B_f\ .
\end{eqnarray}
Here, we have assumed that the enhancement is saturated in the galactic halo, i.e. $\delta$, $\gamma_s \gg v_{\rm rel}^2$. 
Therefore, the self-interaction cross section is much suppressed compared to the annihilation cross section 
in the Breit--Wigner scenario, and hence, dark matter is effectively collisionless. 
This is a drastic difference from the Sommerfeld enhancement scenario where dark matter can have a
large scattering cross section mediated by the light particle.
\\

{\it Effects on Halo Shape}\\
The large self-interaction of the dark matter causes the rapid energy transfer
in the halo and isotropize the velocity dispersion, which leads to a spherical halo and drives the halo towards isothermality. These expectation have been confirmed by simulation in the hard sphere scattering limit~\cite{Spergel:1999mh,Dave:2000ar,Yoshida:2000bx}. The shapes of dark matter halos of elliptical galaxies and clusters are decidedly elliptical, which constraints self-interaction~\cite{MiraldaEscude:2000qt}. 
The elliptical halo constraints for Coulomb interactions have been discussed in Refs.~\cite{Ackerman:2008gi,Feng:2009mn}.

According to Ref.~\cite{Feng:2009hw}, 
we estimate
the impacts of the self-interaction of the enhancement scenarios on the halo shape
by calculating the relaxation time for establishing an isothermal halo. 
Here, we assume the time scale for isotropizing the spatial distribution of the dark matter halo is the same as this relaxation time~\cite{Feng:2009mn}. 
Then, the resultant average rate for dark matter to change velocities by an ${\cal O}(1)$ factor~\cite{Feng:2009hw} is
given by,
\begin{equation}
\Gamma_k=\int d^3v_1 d^3v_2 f(v_1) f(v_2) \left(n_X \vrel \sigma_T \right)
\left(\vrel^2 / v_0^2\right) ,
\end{equation}
where $f(v) = e^{-v^2/v^2_0} / (v_0\sqrt\pi)^3$ is the dark matter's
assumed Maxwellian velocity distribution, $n_X$ is its number density inside the halo.

This rate provides a judgement on the effects of self-interaction in the galactic dynamics. 
If the scattering rate is small and the relaxation time is much longer than the typical age of galaxies,
i.e. $\Gamma^{-1}_k\gg\tau_g\sim10^{10}~{\rm years}$, 
we expect that self-interaction does not play important roles on the galactic dynamics. 
On the other hand, if the scattering is so sufficient and the relaxation time is much shorter than
$\tau_g$, then such scenarios have been excluded by the observed elliptical halos. 
For the parameter region with $\Gamma^{-1}_k$ not far from $\tau_g$, the self-interaction leaves the observable 
imprints on the galaxy's structure.

Now, let us estimate the relaxation time in the Sommerfeld enhancement scenario.
As mentioned earlier, Eq.\,(\ref{eq:sigmatransfer}) receives significant corrections in the strong interaction regime, 
$\beta \gg 1$. In this work, we focus on the $R \gg 1$ region of parameter space. 
In this region, quantum effects are subdominant and hence classical studies of slow moving in plasmas~\cite{Khrapak:2003} are applicable.  
These studies find that numerical analysis of the cross section is accurately reproduced by
\begin{eqnarray}
\sigma_T &\simeq& \frac{4 \pi}{m_\phi^2} \beta^2 \ln \left( 1 +
\beta^{-1} \right)\ , \quad \beta < 0.1 \ , \nonumber \\
\sigma_T &\simeq& \frac{8 \pi}{m_\phi^2} \frac{\beta^2}{1 + 1.5
  \beta^{1.65}} \ , \quad 0.1 < \beta < 1000 \ .
\label{eq:sigmaTclassical}
\end{eqnarray}
We use these fitted cross sections to obtain numerical results given below.

In our analysis,
we consider the well-studied, nearby (about 25 Mpc away) elliptical galaxy NGC 720~\cite{Buote:2002wd,Humphrey:2006rv}. The average dark matter density is $n_X\sim 4~{\rm GeV/cm^3}$ within the 5 kpc where the ellipticity constraint is strong, and the radial velocity dispersion $\overline{v^2_r}(r)\simeq(240~{\rm km/s})^2$~\cite{Feng:2009hw}.

\begin{center}
\begin{figure}[t]
  \includegraphics[width=.85\linewidth]{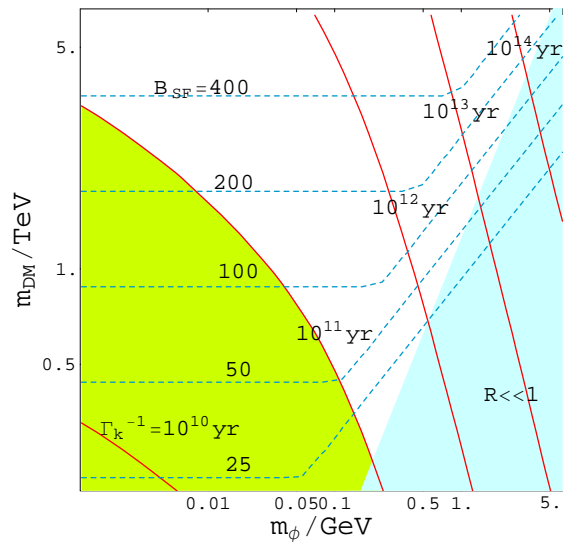}
 \caption{
 Contours of the time-scale of the energy transfer rate (solid lines)
 in the Sommerfeld enhancement scenario.
The dashed lines  show contours of the enhancement factor
obtained with the fine structure constant in Eq.~(\ref{eq:alpha}).
In the light shaded region, the classical cross sections in Eq.~(\ref{eq:sigmaTclassical})
is not applicable.
In the darker shaped region, the relaxation time is not so much longer than the age of galaxy $\tau_g \sim10^{10}$\,years, 
and hence, we expect observable imprints of the self-interaction in the galactic dynamics.
 }
 \label{fig:SF}
\end{figure}
 \end{center}

In Fig.~\ref{fig:SF}, we show the contours of the relaxation time (solid lines) for dark matter changing its energy 
by ${\cal O}(1)$ for a given $m_\phi$ and $m_{\rm DM}$. 
We also plot the contour (dashed lines) for the enhancement factor $B_{SF}\equiv{\rm min}\left[\alpha_X/\vrel,\alpha_X m_{\rm DM}/m_\phi\right]$. In the figure, we have used the fine structure constant determined by the dark matter density\,\cite{Feng:2009hw}, i.e.
\begin{eqnarray}
\label{eq:alpha}
 \alpha_X =\sqrt{\frac{3\times10^{-26}~{\rm cm^3/s}}{\pi}}m_{\rm DM}\simeq 0.029\times \left(\frac{m_{\rm DM}}{1\,{\rm TeV}}\right)\ .
\end{eqnarray}
The figure shows that $\Gamma^{-1}_k$ increases as $m_\phi$ increases
for a given $m_{\rm DM}$. 
This is because for a larger $m_\phi$, the scattering cross section is suppressed. 
For a given $m_\phi$, the larger dark matter mass leads to the larger $\Gamma^{-1}_k$ since 
the number density of dark matter becomes smaller with larger $m_{\rm DM}$, and the scattering rate decreases. 

Notice that in the region with $m_\phi$ smaller than about $100$\,MeV, and the dark matter mass smaller 
than about $3\,{\rm TeV}$, 
the relaxation time scale is less than $10^{11}\,{\rm years}$, which is not so above $\tau_g$.
In this region, the effects of self-interaction are important on the shape of dark matter halos,
while effectively collisionless dark matter has no effects on the halo shape.
Thus, by comparing the predicted halo shapes with observations, we can distinguish the Sommerfeld
enhancement scenario from the other collisionless dark matter scenarios.
To make such comparison precisely, the detailed numerical simulations with velocity-dependent self-interaction are crucial. 
More data sets of NGC~720 and other elliptical galaxies and clusters can make these constraints more robust. 

Dark matter with strong self-interactions also predicts the formation of the constant density cores. 
The time scale for the formation of these cores is again given by $\Gamma^{-1}_k$. In the region with $m_\phi\lesssim100~{\rm MeV}$ and $m_{\rm DM}\lesssim3~{\rm TeV}$, we would expect NGC~720 should have a large core. Future tests for the presence of cores of NGC~720 may provide another way to distinguish the Sommerfeld enhancement scenario from the Breit--Winger enhancement scenario and the other collisionless dark matter model.

Finally, we estimate the relaxation time in the models with the Breit--Winger enhancement.
As we have discussed in the previous section, dark matter is effectively collisionless
in the Breit--Wigner enhancement scenario,
and hence, the energy transfer cross section is much smaller than the enhanced annihilation cross section. For example, we can obtain a desirable boost factor 
for $m_{\rm DM}=1\,{\rm TeV}$, $\delta\simeq 3\times 10^{-5}$ and $\gamma_s\simeq3\times 10^{-5}$
while having the correct dark matter density\,\cite{Ibe:2008ye}.
Then, by using $\sigma_T$ given by Eq.~(\ref{eq:selfBW}), the relaxation time for dark matter to change velocities 
by ${\cal O}(1)$ factors is 
\begin{eqnarray}
\Gamma_k^{-1} \gtrsim 10^{20}\,{\rm years},
\end{eqnarray}
which is much longer than the age of galaxies. 
Here, we have assumed that the inequality in Eq.~(\ref{eq:BWcond1}) is saturated
in the right hand side of the inequality.
Thus, we do not expect that self-interaction in Breit-Wigner enhancement scenario has significant roles in the galactic dynamics.
\\

{\it Summary}\\
The Sommerfeld enhancement and the Breit--Wigner enhancement scenarios were proposed to explain 
electron/positron excesses in cosmic ray. 
Both scenarios can not be distinguished from the cosmic fluxes if we assume certain astrophysical boost factor 
in the Sommerfeld enhancement. 
However, in the Sommerfeld enhancement scenario, the light field, which is essential for the enhancement, 
also mediates strong self-interaction of dark matter. 
The large self-interaction cross section can lead to spherical shapes of dark matter halos.
In the Breit--Wigner scenario, on the other hand, the scattering cross section is much smaller than the enhanced annihilation cross section, and dark matter is effectively collisionless. 
We found that in the Sommerfeld enhancement scenario with 
$m_\phi\lesssim100~{\rm MeV}$ and $m_{\rm DM}\lesssim3~{\rm TeV}$, 
the self-interaction can have measurable effects on the halo shape. 
Such effects can be investigated by careful numerical simulations of the velocity-dependent self-interactions and deeper data sets of the halo shapes.

We comment on the constraints on the enhancement scenarios from the effects on the cosmic microwave 
background (CMB) anisotropy.
As investigated in 
Ref.~\cite{Belikov:2009qx},
the large annihilation cross section of dark matter during and after recombination time results 
in too much energy deposition into background plasma and affects the CMB anisotropy.
The resultant constraints from the CMB anisotropy exclude some portions
of parameter spaces of both the enhancement scenarios.

\section*{Acknowledgements}
The authors appreciate Jonathan Feng and Manoj Kaplinghat for valuable discussions and comments.
M.I. also appreciate T.T~Yanagida for stimulating discussions.
The work of MI was supported by NSF grants PHY--0653656. 
The work of HY was supported in part by NSF grants PHY--0653656 and PHY--0709742.


\begin{thebibliography}{99}
\bibitem{Adriani:2008zr}
  O.~Adriani {\it et al.},
  arXiv:0810.4995 [astro-ph].
\bibitem{:2008zz}
  J.~Chang {\it et al.},
  Nature {\bf 456}, 362 (2008).
\bibitem{Torii:2008xu}
  S.~Torii {\it et al.},
  arXiv:0809.0760 [astro-ph].
\bibitem{Abdo:2009zk}
  A.~A.~Abdo {\it et al.}  [The Fermi LAT Collaboration],
  Phys.\ Rev.\ Lett.\  {\bf 102}, 181101 (2009)
  [arXiv:0905.0025 [astro-ph.HE]].
\bibitem{Hooper:2008kg}
  D.~Hooper, P.~Blasi and P.~D.~Serpico,
  JCAP {\bf 0901}, 025 (2009)
  [arXiv:0810.1527 [astro-ph]];
  H.~Yuksel, M.~D.~Kistler and T.~Stanev,
  Phys.\ Rev.\ Lett.\  {\bf 103}, 051101 (2009)
  [arXiv:0810.2784 [astro-ph]];
  S.~Profumo,
  arXiv:0812.4457 [astro-ph].

\bibitem{Dado:2009ux}
  S.~Dado and A.~Dar,
  arXiv:0903.0165 [astro-ph.HE];
  P.~L.~Biermann \etal,
  Phys.\ Rev.\ Lett.\  {\bf 103}, 061101 (2009)
  [arXiv:0903.4048 [astro-ph.HE]];
  B.~Katz, K.~Blum, E.~Waxman,
  arXiv:0907.1686 [astro-ph.HE].
\bibitem{Komatsu:2008hk}
  E.~Komatsu {\it et al.}  [WMAP Collaboration],
  arXiv:0803.0547 [astro-ph].
\bibitem{Hisano:2004ds}
  J.~Hisano, S.~Matsumoto, M.~M.~Nojiri and O.~Saito,
  Phys.\ Rev.\  D {\bf 71}, 063528 (2005)
  [arXiv:hep-ph/0412403].
\bibitem{ArkaniHamed:2008qn}
  N.~Arkani-Hamed, D.~P.~Finkbeiner, T.~Slatyer and N.~Weiner,
  arXiv:0810.0713 [hep-ph].

\bibitem{Ibe:2008ye}
  M.~Ibe, H.~Murayama and T.~T.~Yanagida,
  Phys.\ Rev.\  D {\bf 79}, 095009 (2009)
  [arXiv:0812.0072 [hep-ph]].
  te{Guo:2009aj}
\bibitem{Guo:2009aj}
  W.~L.~Guo and Y.~L.~Wu,
  Phys.\ Rev.\  D {\bf 79}, 055012 (2009)
  [arXiv:0901.1450 [hep-ph]].



\bibitem{Cirelli:2008pk}
  M.~Cirelli, M.~Kadastik, M.~Raidal and A.~Strumia,
  arXiv:0809.2409 [hep-ph].
\bibitem{Pospelov:2008jd}
  M.~Pospelov and A.~Ritz,
  arXiv:0810.1502 [hep-ph].

\bibitem{Feldman:2008xs}
  D.~Feldman, Z.~Liu and P.~Nath,
  arXiv:0810.5762 [hep-ph].
\bibitem{Griest:1990kh}
  K.~Griest and D.~Seckel,
  Phys.\ Rev.\  D {\bf 43}, 3191 (1991).
  \bibitem{Gondolo:1990dk}
  P.~Gondolo and G.~Gelmini,
  Nucl.\ Phys.\  B {\bf 360}, 145 (1991).
\bibitem{Jungman:1995df}
  G.~Jungman, M.~Kamionkowski and K.~Griest,
  Phys.\ Rept.\  {\bf 267}, 195 (1996)
  [arXiv:hep-ph/9506380].


\bibitem{Feng:2009hw}
  J.~L.~Feng, M.~Kaplinghat and H.~B.~Yu,
  arXiv:0911.0422 [hep-ph].

\bibitem{Buckley:2009in}
  M.~R.~Buckley and P.~J.~Fox,
  arXiv:0911.3898 [hep-ph].

\bibitem{Dent:2009bv}
  J.~B.~Dent, S.~Dutta and R.~J.~Scherrer,
  arXiv:0909.4128 [astro-ph.CO];
  J.~Zavala, M.~Vogelsberger and S.~D.~M.~White,
  arXiv:0910.5221 [astro-ph.CO].

\bibitem{Ibe:2009en}
  M.~Ibe, H.~Murayama, S.~Shirai and T.~T.~Yanagida,
  arXiv:0908.3530 [hep-ph].


\bibitem{Dave:2000ar}
  R.~Dave, D.~N.~Spergel, P.~J.~Steinhardt, B.~D.~Wandelt,
  Astrophys.\ J.\  {\bf 547} (2001) 574
  [arXiv:astro-ph/0006218].

\bibitem{Yoshida:2000bx}
  N.~Yoshida, V.~Springel, S.~D.~M.~White, G.~Tormen,
  Astrophys.\ J.\  {\bf 535}, L103 (2000)
  [arXiv:astro-ph/0002362];
  B.~Moore \etal,
  Astrophys.\ J.\  {\bf 535}, L21 (2000)
  [arXiv:astro-ph/0002308];
  M.~W.~Craig, M.~Davis,
  arXiv:astro-ph/0106542;
  C.~S.~Kochanek, M.~J.~White,
  Astrophys.\ J.\  {\bf 543}, 514 (2000)
  [arXiv:astro-ph/0003483].

\bibitem{Spergel:1999mh}
  D.~N.~Spergel and P.~J.~Steinhardt,
  Phys.\ Rev.\ Lett.\  {\bf 84} (2000) 3760
  [arXiv:astro-ph/9909386].

\bibitem{MiraldaEscude:2000qt}
  J.~Miralda-Escude,
  arXiv:astro-ph/0002050.

\bibitem{Ackerman:2008gi}
  L.~Ackerman, M.~R.~Buckley, S.~M.~Carroll and M.~Kamionkowski,
  Phys.\ Rev.\  D {\bf 79}, 023519 (2009)
  [arXiv:0810.5126 [hep-ph]].

\bibitem{Feng:2009mn}
  J.~L.~Feng, M.~Kaplinghat, H.~Tu and H.~B.~Yu,
  JCAP {\bf 0907}, 004 (2009)
  [arXiv:0905.3039 [hep-ph]].

\bibitem{Khrapak:2003}
S.~A.~Khrapak \etal,
Phys.\ Rev.\ Lett.\  {\bf 90}, 225002 (2003);
IEEE Transactions on Plasma Science {\bf 32}, 555 (2004).

\bibitem{Buote:2002wd}
  D.~A.~Buote, T.~E.~Jeltema, C.~R.~Canizares and G.~P.~Garmire,
  Astrophys.\ J.\  {\bf 577} (2002) 183
  [arXiv:astro-ph/0205469].

\bibitem{Humphrey:2006rv}
P.~J.~Humphrey \etal,
  Astrophys.\ J.\  {\bf 646} (2006) 899
  [arXiv:astro-ph/0601301].

\bibitem{Belikov:2009qx}
  A.~V.~Belikov and D.~Hooper,
  Phys.\ Rev.\  D {\bf 80}, 035007 (2009)
  [arXiv:0904.1210 [hep-ph]];
  S.~Galli, F.~Iocco, G.~Bertone and A.~Melchiorri,
  Phys.\ Rev.\  D {\bf 80}, 023505 (2009)
  [arXiv:0905.0003 [astro-ph.CO]];
  G.~Huetsi, A.~Hektor and M.~Raidal,
  arXiv:0906.4550 [astro-ph.CO];
  M.~Cirelli, F.~Iocco and P.~Panci,
  JCAP {\bf 0910}, 009 (2009)
  [arXiv:0907.0719 [astro-ph.CO]];
  T.~R.~Slatyer, N.~Padmanabhan and D.~P.~Finkbeiner,
  Phys.\ Rev.\  D {\bf 80}, 043526 (2009)
  [arXiv:0906.1197 [astro-ph.CO]];
  T.~Kanzaki, M.~Kawasaki and K.~Nakayama,
  arXiv:0907.3985 [astro-ph.CO].


\end{thebibliography}
\end{document}